\documentclass[longauthor]{aastex62}

\received{2019 April 17}
\revised{2019 June 9}
\accepted{2019 June 9}
\submitjournal{ApJL}

\shorttitle{A non-ejecting thermonuclear runaway}
\shortauthors{Hillman et al.}

\begin{document}

\title{The supersoft X-ray transient ASASSN-16oh as a thermonuclear
 runaway without mass ejection}

\correspondingauthor{Yael Hillman}
\email{yaelhi@ariel.ac.il}

\author{Yael Hillman}
\affil{Department of Physics, Ariel University, Ariel POB 3, 40700, Israel}

\author{Marina Orio}
\affiliation{Department of Astronomy, University of Wisconsin
475 N. Charter Str.  Madison, WI 53706}
\affiliation{INAF - Astronomical Observatory Padova, vicolo dell'Osservatorio 5, 35122 Padova, Italy}

\author{Dina Prialnik}
\affil{Department of Geosciences, Tel Aviv University,\\
 Ramat Aviv, Tel Aviv 69978, Israel}

\author{Michael Shara}
\affil{Department of Astrophysics, American Museum of Natural History,\\
 Central Park West and 79th Street, New York, NY 10024-5192, USA}

\author{Pavol Bez{\'a}k}
\affil{Advanced Technologies Research Institute, Faculty of Materials Science and Technology in Trnava, Slovak University of Technology in Bratislava, Bottova 25, 917 24 Trnava, Slovakia}

\author{Andrej Dobrotka}
\affiliation{Advanced Technologies Research Institute, Faculty of Materials Science and Technology in Trnava, Slovak University of Technology in Bratislava, Bottova 25, 917 24 Trnava, Slovakia}

\begin{abstract}
 The supersoft X-ray and optical transient ASASSN-16oh
has been interpreted by \citet{maccarone2019}
as having being induced by an accretion event on a massive white dwarf, resembling a dwarf nova super-outburst.
 These authors argued that the supersoft
 X-ray spectrum had a different origin than in 
 an atmosphere heated by shell nuclear burning, because no mass was ejected.
We find instead that the event's timescale and other characteristics 
 are typical of non-mass ejecting thermonuclear runaways, as already
 predicted by \cite{shara1977} and the extensive grid of nova models by \citet{yaron2005}. 
 We suggest that
 the low X-ray and bolometric luminosity in comparison to
 the predictions of the models of nuclear burning are due 
 to  an optically 
 thick accretion disk, hiding most of the white dwarf surface. 
 If this is the case, we
 calculated that the optical transient can be explained as a non-ejective thermonuclear event on a WD of $\simeq$1.1M$_\odot$ accreting at the rate of $\simeq3.5{-}5{\times}10^{-7}$M$_\odot$yr$^{-1}$. We make predictions that should prove whether the
 nature of the transient event was due to thermonuclear burning or to
 accretion; observational proof should be obtained in the next few years,
because a new outburst should occur
 within $\simeq$10-15 years of the event.  
\end{abstract}


\section{Introduction} \label{sec:intro}
ASASSN-16oh was discovered on December 2 2016 by the All-Sky Automated
Survey for Supernovae, as a V=16.9 transient in the field of the
 Small Magellanic Cloud (SMC) \citep{shappee2014}. It was then observed
 with the Neil Gehrels Swift
Observatory, whose X-ray telescope showed that on 2016 December 15
a supersoft X-ray source (hereafter, SSS) was the X-ray
 counterpart of the optical transient, with a blackbody temperature of about
900,000K and a luminosity of about  10$^{37}$ erg s$^{-1}$ (assuming the
source is located in the SMC, \citet{maccarone2019}). The optical spectra showed narrow optical
emission lines (unresolved with a resolution of about 300 km s$^{-1}$)
at a velocity consistent with the systemic one of the SMC \citep{maccarone2019}.

 OGLE data showed that the optical magnitude decreased from I$\simeq$21 
 to a peak at I=16.5 within an interval of 296 days
 (although the rise could not be monitored for a period of over
 4 months at the beginning, so there is some uncertainty),
 and a decay to I$\simeq$19.5
 followed in an interval of between 259 and 268 days after maximum brightness \citep[see][]{maccarone2019}.
 There was a successive brightening to   
 magnitude I$\simeq$19 and a decay to I$\simeq$21 with another oscillation
 \citep{maccarone2019}, but it is not clear whether the oscillations in luminosity in the last part of the light curve are
 manifestations of the return to minimum or part of a pattern observed
 by OGLE in the 6 years before the outburst, when the magnitude
 oscillated between I$\simeq$21 and I$\simeq$19.5 at least four times.

 \citet{maccarone2019} have argued that this event cannot be due to
 thermonuclear burning, which is the cause of the vast majority
 of SSS in the Galaxy and in the Magellanic Clouds, because no
 mass was ejected, and the rise to maximum was much longer and the amplitude lower than in novae.
 Thus, they proposed a qualitative model of a peculiar disk instability.
 These authors suggested that a superoutburst 
 in a cataclysmic variable with a massive WD (1.2 M$_\odot$) accreting at
 about 3 $\times 10^{-7}$ M$_\odot$ year$^{-1}$ produces
 the observed supersoft X-ray flux because a ``spreading
 layer'' is suddenly accreted onto the WD surface, dissipating
 the rotational energy of the material incoming
 from the accretion disk  \citep[]{kippenhahn1978,
 piro2004}.
 Their underlying idea is that the disk instability in this transient occurred before the conditions for triggering nuclear burning have had a chance to develop, thus there was no ignition to lead to an ejected shell.

 The theoretical models, however, have never 
 predicted that a thermonuclear runaway on a WD always leads to mass
 ejection.  Quite the contrary,
 since \citet{shara1977} and \citet{fujimoto1982}, non-ejecting thermonuclear
 runaway events \textit{have} been predicted. Quantitative predictions of such
 events were published in detail by \citet{yaron2005}. As the mass
 transfer rate onto the WD increases, there is no abrupt transition 
 to the regime in which all the generated energy is radiated (that
 of {\it steady supersoft X-ray sources}). There is nearly an order of magnitude range of $\dot{M}$ values wherein the outburst causes only a slow increase in
 luminosity, followed by a slow decay.  At a very high WD mass 
 of 1.3 M$_\odot$,  the 
 rise and decay can be short, just tens of days. In contrast, on a 0.6 M$_\odot$
 WD the time scales are tens of thousands of days \citep[see][]{yaron2005}.

 The only observational fact that casts serious doubt on the thermonuclear
 outburst hypothesis in ASASSN-16oh is the low bolometric luminosity,
 between 10$^{35}$ erg s$^{-1}$ and 10$^{36}$ erg s$^{-1}$.
 \citet{maccarone2019} find a luminosity of the order of 10$^{36}$ erg s$^{-1}$
 with a blackbody fit, and about the same
 value  by fitting an atmospheric
 model. We note that
 the {\sl Chandra} LETG spectrum has poor S/N and the absorption features cannot
 be measured for proper model-fitting, so the atmospheric models cannot
 be ``pegged''  using the absorption features.

 Thermonuclear runaways, regardless of the mass ejection, cause the
 WD to radiate at Eddington luminosity, around 10$^{38}$ erg s$^{-1}$.
 However, many novae have been found to be SSS with much lower  
 luminosity than the theoretical prediction, often
 only 1/100th of Eddington luminosity. This is not thought to occur,
 in most cases, 
 because the WD is obscured by non-isotropically
 expanding ejecta,
 although this may be the case of some novae {\citep{beardmore2010,beardmore2012,tofflemire2013}.
Instead, a non-disrupted, or quickly reformed, accretion disk seems to 
 block all or part of the flux from the central source for systems
 at high inclination \citep{ness2013}. In some cases, as in U Sco,
the SSS is still
 observed even if the WD is completely blocked, because of Thomson
 scattering \citep{ness2012, ness2013, orio2013}.
 \citet{maccarone2019} discussed the emission lines 
 optical spectrum, and found that the lines
 are very narrow because the accretion
 disk must be fully ionized, so they are produced only in the 
outermost portion of the disk. 
\section{The Chandra observation}
ASASSN-16oh was observed by PI Maccarone with {\sl Chandra} and the HRC-S camera
 and LETG grating for 50 kiloseconds on 2017 March 29. The data were
 reported in \citet{maccarone2019}. We downloaded and analyzed them in
order to fit the spectrum with a WD atmospheric model instead of a blackbody
\citep[shown by][]{maccarone2019},
 and to perform a timing analysis. We extracted the {\sl Chandra} HRC-S+LETG with
 their first order grating redistribution matrix files and ancillary
response files with the CIAO 4.9 task chandra repro, with the 4.7 version of the calibration package CALDB. 
 Our best fit is obtained with  a ``halo'' (metal poor) model calculated with
 TMAP (from the web site \url{http://astro.uni-tuebingen.de/#rauch/TMAP/TMAP.html}),
 the physics is
 described in detail in  \citet{werner2003, rauch2010}. 
We
coadded the positive and negative order spectra with {\it combine grating 
spectra} to increase the signal-to-noise ratio.

As noted by \citet{maccarone2019}, an
 atmospheric model with less-than-solar 
abundances, appropriate for the Galactic halo, fits the data better than
 a model with enhanced abundances, and is more appropriate for SSS in the SMC \citep[e.g.][]{orio2018}.
  As an alternative to the blackbody model
 plotted by the above authors in the first figure of their
 article, we present this atmospheric model in  Fig.1. 
 Given the data quality, this
 fit is statistically as acceptable as the blackbody.
 Because the signal-to-noise 
 is poor, like \citet{maccarone2019} we used C statistics \citep{cash1979} obtaining C-statistic parameter
1716 for 1710 degrees of freedom in the range 10-70 \AA. The best fit shown in Fig.1 is 
obtained with a temperature T$_{\rm eff}$ of 750,000K, a  column
density of N(H)=2.3 $\times 10^{20}$ cm$^{-2}$, an unabsorbed flux
in the 0.1-1 keV X-ray band of 5.3 $\times 10^{-12}$ erg cm$^{-2}$ s$^{-1}$ and
a bolometric luminosity at the SMC distance of  
4.3 $\times 10^{36}$ erg s$^{-1}$. 
Because the best-fit column density appears low for the SMC,
by constraining the value of N(H) to
N(H)=4 $\times 10^{20}$ cm$^{-2}$ we obtain with 
T$_{\rm eff}$=736,000K the same unabsorbed flux 
and a bolometric luminosity at the SMC distance of
6.4 $\times 10^{36}$ erg s$^{-1}$. This fit with 
higher N(H) is still statistically
acceptable, and given the data quality it cannot be ruled out.   

\begin{figure}
\begin{center}
\resizebox{0.8\hsize}{!}{\includegraphics{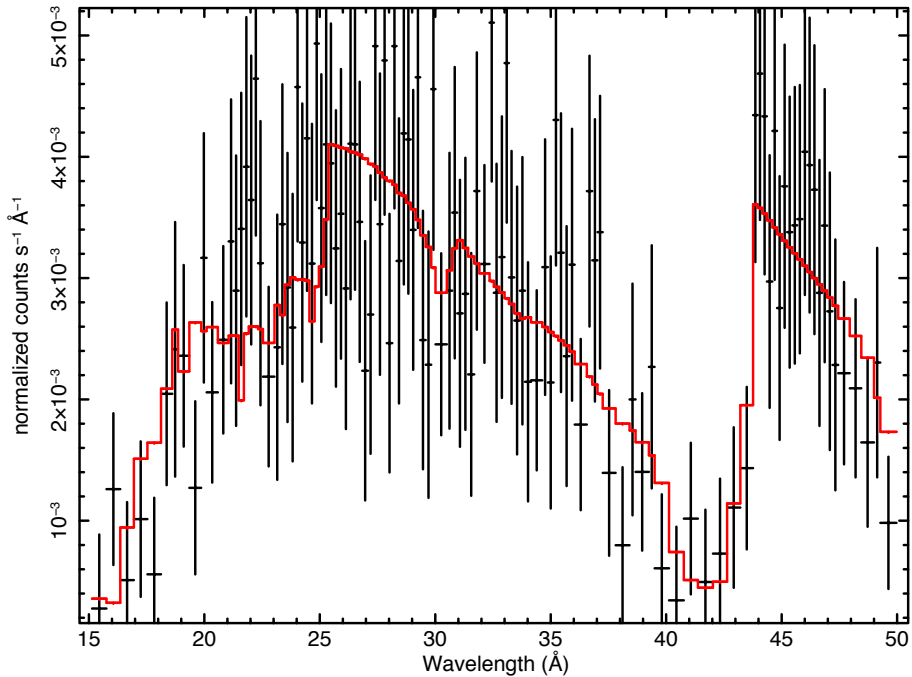}}
\end{center}
\vspace{-0.7cm}
\caption{Best fit to the {\sl Chandra} spectrum with 
 a NLTE atmospheric model with T$_{\rm eff}=750,000$K, 
 N(H)=2.3 $\times 10^{20}$ cm$^{-2}$, unabsorbed flux
 in the 0.1-1 keV X-ray band of 5.3 $\times 10^{-12}$ erg cm$^{-2}$ s$^{-1}$ and
 a bolometric luminosity at the SMC distance of
4.3 $\times 10^{36}$ erg s$^{-1}$. 
}
\end{figure}
\begin{figure}
\begin{center}
\resizebox{0.9\hsize}{!}{\includegraphics{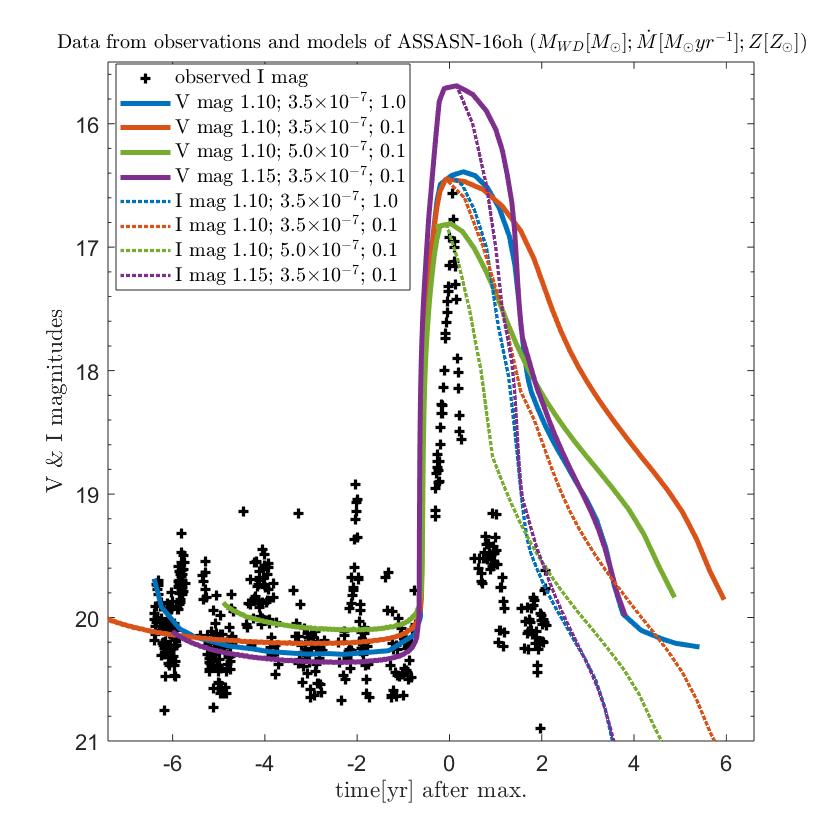}}
\end{center}
\vspace{-0.7cm}
\caption{The observed I band of the light curve of ASASSN-16oh (black plus signs) compared with the predicted luminosity of the non-ejecting nova model in the V band (solid curves) and the I band (dashed curves) of four models: $M_{WD}{=}1.1M_\odot$, $\dot{M}{=}3.5{\times}10^{-7}M_\odot yr^{-1}$, solar metalicity (blue); $M_{WD}{=}1.1M_\odot$, $\dot{M}{=}3.5{\times}10^{-7}M_\odot yr^{-1}$, one tenth of solar metalicity (red); $M_{WD}{=}1.1M_\odot$, $\dot{M}{=}5{\times}10^{-7}M_\odot yr^{-1}$, one tenth of solar metalicity (green); $M_{WD}{=}1.15M_\odot$, $\dot{M}{=}3.5{\times}10^{-7}M_\odot yr^{-1}$, one tenth of solar metalicity (purple).}
\end{figure}
%
%
\section{A non-ejecting nova model}
 The rise to maximum in the case of a thermonuclear runaway that
does not eject mass is very long 
 compared to a runaway that
produces a classical nova event without mass ejection.
 The duration of a thermonuclear event
 is dictated by the expansion velocity {\textemdash} the faster the
 envelope expands, the faster the maximum luminosity is attained.
  The velocity varies inversely with the accretion rate,  i.e.,
 a slow accretion rate yields a fast velocity, while faster rates yield
 slower velocities \citep[Table 3]{yaron2005}. At extremely high accretion
 rates, the velocity of the mass in the outer shells of the WD's envelope
 becomes so slow that it does not exceed the WD escape velocity,
 resulting in a slow expansion and contraction, entirely without mass ejection.

Our models are produced by using a hydrodynamic Lagrangian nova evolution code, that simulates the evolution of multiple consecutive cycles of a WD accreting mass, the triggering of a TNR, and the 
 onset of the outburst. Further details regarding the code may be found in \cite{prialnik1995,epelstain2007,Hillman2015}.  

Turning to the extensive grid of models by \cite{yaron2005}, we find that the models that best fit the data are those of a  1.1M$_\odot$ WD accreting at $3.5{-}5{\times}10^{-7}$ M$_\odot$ yr$^{-1}$. The recurrence  time of the thermonuclear outburst is ${\sim}10$ years, and each V band outburst lasts almost 4 years. A comparison with the observed light curve is shown in Fig.2. The maximum effective temperature (T$_{\rm eff}$) the model reaches is ${\sim}750,000$K, which is consistent with the value we obtain 
 by fitting the {\sl Chandra} data. 

 The accreted matter of this model, as for the entire grid of models by \cite{yaron2005}, is of solar metalicity. The metalicity in the SMC is about one tenth of the solar metalicity. We therefore carried out three additional simulations, each one varying from the above model by either mass, accretion rate or metalicity, in order to understand how sensitive the results are to changes in these parameters. The light curves for these models, in the V and I bands, are presented in Fig.2. We find that lowering the metalicity, while not changing the mass and accretion rate lengthens the outburst duration by ${\sim50\%}$. 
 In addition, we find that decreasing the accretion rate, while not changing the WD mass or metalicity, increases the outburst amplitude and duration, as we see for the solar metalicity grid of models. 
  Decreasing only the WD mass at constant accretion rate and metalicity decreases the outburst amplitude and increases its \textbf{duration}, also in agreement with the behavior of the grid of solar metalicity models. We conclude, that changing the metalicity of the accreted matter will produce similar behavioral trends 
  to what we see in the grid, with somewhat of a shift in the initial input parameters (i.e. $M_{WD}$ and $\dot{M}$), the general regime of the results remaining as discussed here and demonstrated in Fig.2. 
 
Our code produces the V band light curve, while the OGLE data is in the I band. We show these V band light curves as a prediction of the light curve behavior in this band, which can be tested during the next eruption, provided it will be observed in the V band. The V band light curves in Fig.2 exhibit a longer outburst duration than that of the observed I band. This discrepancy between the V and I bands is due to the effective temperature in TNR events rising quickly, shifting the light curve towards faster frequencies, i.e., away from the IR. Therefore, the I band will decline earlier than the V band. The dashed curves in Fig.2 show an estimate of the I band of our models We produced this estimate based on the computed median of the V-I mag of over 30 classical novae light curves from \citet[fig.8a]{shara2016} showing a gradual decline from V-I$\simeq$0 to V-I$\simeq$1 over $\sim$40 days, and on the longer term V-I behaviors of IM Nor and CI Aql \cite[table 25]{schaefer2010} showing a decline to  V-I$\simeq$1.3 a few years after maximum. Fig.2 demonstrates how the estimated I band decline, for all four models, begins earlier than that of the V band, bearing a good resemblance with the dimming time scale of the OGLE data.

 A special characteristic of the OGLE light curve obtained with
 over 8 years of monitoring, of which 6 are before the beginning of the
 outburst, are the small scale flares ($\approx$1 mag oscillations)
 observed 6 times, approximately once a year. 
  The duration of these flares would be unusually long for small
 amplitude dwarf nova outbursts, on the other hand phenomena of irregular variability with a $\simeq$1 mag amplitude,
 possibly different from the dwarf nova phenomenon,  are not unusual in classical and recurrent novae and a few examples of pre- and
 post-outburst fluctuations can be found in \citet{collazzi2009}. 
\section{Timing analysis of the Chandra light curve}
We examined the {\sl Chandra} zero-order light
curves measured with the HRC-S camera
 in the exposure described in Section 2, \citet{maccarone2019}, because thermonuclear burning SSS
seem to be associated either with
 phenomena of short pulsations lasting around a minute or less \citep[see][and references therein]{ness2015}, or with longer 
periodicities of the order of half an hour. For the latter, examples are V1494
 Aql \citep{drake2003}, V4743 Sgr \citep{leibowitz2006, zemko2016,
dobrotka2017}, V2491 Cyg \citep{ness2011, zemko2015}, V959 Mon \citep{peretz2016}.
These periodic modulations over tens of minutes have been attributed to
the WD spin for V4743 Sgr and V2491 Cyg. 
 A possible explanation for the SSS in those novae is that the WD is magnetic, namely an {\it intermediate
 polar.} During burning, the surface of the
 magnetic caps, heated also from above by accretion, may be at a higher
 temperature than the rest of the WD surface. This may have been the case of the intermediate polar nova V407 Lup, observed with {\sl Chandra} \citep{aydi2018}. Another interesting possibility proposed for V407 Lup and other SSS is that the burning was confined to the magnetic polar regions for a certain period of time \citep{orio1993,king2002,aydi2018}. 

In the Lombe-Scargle periodograms of the {\sl Chandra} data
 of our source, performed after a polynomial detrending to take possible irregular variability into account, we found two peaks at 0.6 and 0.8 mHz below the 70\%
 confidence level, and no peak reaching the 90\% confidence level. However, in the first half of the exposure,
the 0.8 mHz signal, corresponding  to about 20.83 minutes,
is at the 86\% confidence level. Although this is not a high confidence
 detection, the modulation may be real and may not have been measured in
 the second half of the exposure because its amplitude decreased. 
 Such changes in the short term modulations of the soft X-ray flux over the time of an exposure seem to be common in novae \citep{ness2011,dobrotka2017}. 
\section{Discussion \& Conclusions}
 Modeling shows that thermonuclear runaway events on WD's (i.e., nova events) with a lack of (or very little) mass ejection are possible, providing the mass accretion rate is sufficiently high, at
 least a few times $10^{-7}M_\odot yr^{-1}$ \cite[]{prialnik1995,yaron2005,starrfield2012,hillman2016}. 

When fitting the {\sl Chandra} data with an atmospheric model, we tested models with
enhanced and metal-poor abundances. The best fit was obtained
with the metal-poor ``halo'' model, which is also
consistent with SMC abundances. 

We have shown that the time scale of quiescence (accretion), and of the rise and decline (eruption) are in excellent agreement with our nova model of a WD mass of $1.1M_\odot$ and accretion of solar metalicity matter at a high accretion rate of $3.5{-}5{\times}10^{-7}M_\odot yr^{-1}$, while the metal-poor examples that we have presented here are also in good agreement with the light curve. Examining the four models demonstrates the impact of altering the different parameters will have on the eruptive behavior, in particular, it shows that 
 the change caused by altering the metalicity can be compensated for by tweaking one (or both) of the other parameters.

 The effective temperature of the fit is in excellent agreement with the maximum $T\rm_{eff}$ produced by the model. The variance of the modeled $T\rm_{eff}$ is small compared with a mass ejecting eruption, it remains above $\sim{400,000}$K throughout the event. This is because the TNR is relatively weak so the WD radius expands less than in a mass ejecting event. The maximum modeled radius is always $<{0.1M_\odot}$ so the cooling of the WD's photosphere is very moderate. This means the WD is always an SSS during TNR of this class of non-ejecting events, which is in agreement with the  {\sl Swift} observations that overlap the time of the observed OGLE I mag. 
The observed brightness is at least two orders of magnitude lower
 than the modeled brightness. 
However, the combination of observed luminosity and effective temperature would
yield a very unlikely, nearly Chandrasekhar mass WD.
 We interpret the discrepancy in observed and modeled luminosity
 as due to obscuration,  an explanation already brought forward
 in recent years
 to explain the low SSS luminosity in several novae as due to
 the accretion disk in a high inclination binary.  
 There is no information on the inclination of this system,
 because the accretion disk
 is fully ionized, as discussed in the Introduction.
 At the high mass accretion rate we inferred, the disk
 should remain fully ionized even at quiescence, so
 determining the inclination from the emission line profiles
  does not seem feasible even after the outburst. We note
 that the disk in outburst must have
 been very flared and very optically thick,
 consistent with the  {\sl Swift} UVOT
 light curves published by \citet{Rajoeli2017}, which indicate 
 UV luminosity higher than the I luminosity by a factor of
 almost 10. This large UV luminosity must arise in the disk, because
 with a peak temperature of 750,000 K the observed UV flux 
 is too high to be due  the Raleigh-Jeans tail of the SSS.
 Thus, regardless of the inclination, it is very likely that the accretion
 disk was optically thick to supersoft X-rays,
 geometrically thick  and completely
 opaque to the SSS flux coming from most of the WD surface.
 We suggest that this explains the low measured bolometric
 luminosity compared
 with the thermonuclear runaway model.   

Indeed, novae are frequently reported to exhibit an observed brightness that is merely a fraction of the true luminosity. For example, due to the high inclination angle of U Sco, the WD is not directly observed. For this object it is flux from Thomson scattering that is measured, which conserves the WD spectral shape and features, but at a lower intensity than the true WD luminosity. For U Sco it is $\sim 10\%$ \cite[]{orio2013}.
\citet{ness2013} discuss the cases of three other novae with high inclination,
V1494 Aql, V959 Mon, and HV Cet, in which the SSS luminosity is
 only partially detected. V959 Mon was analyzed in detail by \citet{peretz2016}.
The ``missing SSS flux'' is also characteristic of two persistent
 SSS's, CAL 87 
 \citep{orio2004,ness2013} and QR And 
 \citep{ness2013}, which are also high inclination objects. 

\cite{maccarone2019} suggested
 that ASASSN-16oh cannot be a nova eruption due to the lack of an observed shell, meaning that the WD did not eject mass. We have shown via modeling that the lack of ejected mass does not indicate 
that there is no thermonuclear burning, but rather that the
 expansion velocity is slower than the escape velocity from the WD surface. This is caused by a very high mass transfer rate, which minimizes degeneracy and hence explosivity in the accreted envelope of ASSASN-16oh. The resulting recurrent thermonuclear runaway is nonejective.
 
One SSS in M31, CXO J004318.8+412016, is estimated to be a WD with a mass of $\geq1.2M_\odot$ and an accretion rate $>10^{-8}M_\odot yr^{-1}$ and has been reported to be consistent with a post-thermonuclear
 outburst X-ray behavior of a very rapidly recurring source, 
possibly with a time of only a few months.
 No optical outburst has been observed from
 this source, implying no mass loss \cite[]{orio2017}.
This may be the case for other recurring SSS observed in M31 \citep{orio2010}
and in external galaxies \citep[see catalog by][]{wang2016}.

We have shown that the characteristics of ASASSN-16oh are typical of a non-ejective thermonuclear
 outburst, making it highly plausible that this object is indeed a non-ejecting nova. 
The models discussed here have recurrence periods of order ten to fifteen years, leading to the prediction that ASASSN-16oh may erupt again in about a decade,
 supporting the hypothesis that it is a typical recurrent thermonuclear
 runaway event. 
\bibliographystyle{aasjournal}


\end{document}